\documentclass[aps,pra,notitlepage,twocolumn,showpacs,superscriptaddress]{revtex4-1}
\usepackage{amsfonts}
\usepackage{}
\usepackage{amssymb}  
\usepackage{blindtext}
\usepackage{graphicx}  
\usepackage{dcolumn}   
\usepackage{amsmath,bm,amssymb,latexsym,galois,amsthm}   
\usepackage{dsfont} 
\usepackage{mathrsfs}
\usepackage[all,cmtip]{xy}
\usepackage{bm}

\usepackage{subfigure}
\usepackage{rotating}

\usepackage{ifpdf,xy}
\usepackage[braket]{qcircuit}
\usepackage{epstopdf}
\usepackage{graphicx}

\def\bn{\begin{definition}}
\def\en{\end{definition}}
\def\ba{\begin{array}}
\def\ea{\end{array}}
\def\be{\begin{equation}}
\def\ee{\end{equation}}
\def\bd{\begin{description}}
\def\ed{\end{description}}
\def\bu{\begin{enumerate}}
\def\eu{\end{enumerate}}
\def\bi{\begin{itemize}}
\def\ei{\end{itemize}}
\usepackage[colorlinks,
            linkcolor=red,
            anchorcolor=red,
            citecolor=red,
            ]{hyperref}

\theoremstyle{plain}
\theoremstyle{definition}
\newtheorem{definition}{Definition}
\newtheorem{theorem}{Theorem}
\newtheorem{lemma}{Lemma}

\newtheorem*{defn*}{Definition}
\def\eproof{{\mbox{}\hfill\qed}\medskip}

\theoremstyle{remark}

\newcommand{\Z}{{\mathbb{Z}}}
\newcommand{\R}{{\mathbb{R}}}
\newcommand{\C}{{\mathbb{C}}}

\begin{document}

\title{\bf State preparation based on quantum phase estimation}
\author{Jian Zhao}
\affiliation{Key Laboratory of Quantum Information, Chinese Academy of Sciences, School of Physics, University of Science and Technology of China, Hefei, Anhui, 230026, P. R. China}
\affiliation{CAS Center For Excellence in Quantum Information and Quantum Physics, University of Science and Technology of China, Hefei, Anhui, 230026, P. R. China}
\author{Yu-Chun Wu}
\email{Email address: wuyuchun@ustc.edu.cn}
\affiliation{Key Laboratory of Quantum Information, Chinese Academy of Sciences, School of Physics, University of Science and Technology of China, Hefei, Anhui, 230026, P. R. China}
\affiliation{CAS Center For Excellence in Quantum Information and Quantum Physics, University of Science and Technology of China, Hefei, Anhui, 230026, P. R. China}
\author{Guang-Can Guo}
\affiliation{Key Laboratory of Quantum Information, Chinese Academy of Sciences, School of Physics, University of Science and Technology of China, Hefei, Anhui, 230026, P. R. China}
\affiliation{CAS Center For Excellence in Quantum Information and Quantum Physics, University of Science and Technology of China, Hefei, Anhui, 230026, P. R. China}

\author{Guo-Ping Guo}
\email{Email address:  gpguo@ustc.edu.cn}
\affiliation{Key Laboratory of Quantum Information, Chinese Academy of Sciences, School of Physics, University of Science and Technology of China, Hefei, Anhui, 230026, P. R. China}
\affiliation{CAS Center For Excellence in Quantum Information and Quantum Physics, University of Science and Technology of China, Hefei, Anhui, 230026, P. R. China}
\affiliation{Origin Quantum Computing Hefei, Anhui 230026, P. R. China}

\begin{abstract}
State preparation is a process encoding the classical data into the quantum systems. Based on quantum phase estimation, we propose the specific quantum circuits for a deterministic state preparation algorithm and a probabilistic state preparation algorithm. To discuss the gate complexity in these algorithms, we decompose the diagonal unitary operators included in the phase estimation algorithms into the basic gates. Thus, we associate the state preparation problem with the decomposition problem of the diagonal unitary operators. We analyse the fidelities in the two algorithms and discuss the success probability in the probabilistic algorithm. In this case, we explain that the efficient decomposition of the corresponding diagonal unitary operators is the sufficient condition for state preparation problems.
\end{abstract}
\maketitle

\section{Introduction}

When processing machine learning tasks, it is reasonable to presume that the potential effect of quantum computers may outperform classical computers \cite{biamonte2017quantum}. Within quantum machine learning, it is necessary to map the classical data to the quantum systems (e.g. the quantum states or the quantum transformations). This encoding process is called state preparation \cite{schuld2018supervised}.
It is used widely in the quantum algorithms, such as quantum principal component analysis \cite{lloyd2014quantum} and the HHL algorithm for linear systems \cite{harrow2009quantum}.

Given a classical real (complex) vector data set $X=\{x^i\}\subset \R^N(\C^N)$ as the input, our goal is to encode the classical data to the quantum states or the quantum transformations by keeping to some rules.
According to the encoding approaches, the methods for state preparation include basis encoding, qsample encoding, hamiltonian encoding and amplitude encoding.
The basis encoding has been introduced to prepare the superposition state of computing basis for binary strings \cite{ventura2000quantum}; the qsample encoding is to associate a real amplitude vector  with a classical discrete probability distribution \cite{andrieu2003introduction,schuld2018supervised}; the hamiltonian encoding is to focus a Hamiltonian of a system with a matrix that represents the data \cite{lloyd1996universal,georgescu2014quantum};  the amplitude encoding means to encode the data set into the amplitudes of a quantum state, the details seen in Ref. \cite{schuld2018supervised}.
In the amplitude encoding, for each input $x^i\in\R^N(\C^N)$ it needs $log_2{N}$ qubits to acquire the input amplitude features \cite{grover2002creating,kaye2004quantum,mottonen2004transformation,soklakov2006efficient,giovannetti2008quantum}.

We mainly discuss the amplitude encoding methods. It seems hard to achieve exponential speedup for loading $N$ features. Indeed the gates complexity in quantum circuits for state preparation problems is at least $\mathcal O(N)$ in general case or the worst case. Thus the depth of quantum circuits is larger than $\mathcal O(\log_2 N)$ when the width of the circuits is limited to $\mathcal O(log_2 N)$, which means in this case the runtime of state preparation is not efficient.
Cortese and Braje present an amplitude encoding method where the classical data is a binary strings set and the circuits depth  reaches $\mathcal O(log_2 N)$ since the scale of input qubits is $\mathcal O(N)$ \cite{cortese2018loading}. Another noteworthy topic is to make a trade-off between circuits width and depth. One might construct an amplitude encoding algorithm with $\mathcal O(N)$ ancillary qubits in $\mathcal O (n)$ runtime. And generally it is still a significant open question which classic data sets can be efficiently prepared. In this paper, based on phase estimation method, we attribute the state preparation problem to the decomposition problem of the diagonal unitary operators. We propose the specific quantum circuits for a deterministic state preparation algorithm inspired by Ref. \cite{grover2002creating,kaye2004quantum} and a probabilistic state preparation algorithm. We will explain that the efficient decomposition of the corresponding diagonal unitary operators is the sufficient condition for state preparation.

The article is organized as follows. At the end of this section we briefly state the notations used in this paper.
In section \ref{sec_main}, we describe two kind algorithms represented by quantum circuits. The deterministic algorithm and the the probabilistic algorithm for state preparation are presented in section \ref{subsec_dtm} and section \ref{subsec_prb}. The key diagonal unitary operators are contained in the quantum circuits for state preparation. In section \ref{subsec_diag}, we describe the decomposition of diagonal unitary operators with a certain precision.
We leave the discussion and quantitative analysis of the algorithms in section \ref{sec_discuss}.
We give quantitative estimations of success probability and fidelity theoretically.
At last in section \ref{sec_conclusion}, we draw the conclusions of this paper.

\textbf{Notation.}
We use capital Roman letters $A$, $B$,$\ldots$, for matrices , lower case Roman letters $x$, $y$,$\ldots$, for column vectors,
and Greek letters $\alpha$, $\beta$,$\ldots$, for scalars.
Given a column vector $x$, $x^T$ denotes its transpose and
$x^{\dag}\triangleq(\bar x)^T$ is its conjugate transpose, and similar for a given matrix $A$.
Specifically, for the unitary transformation $U$, $U^{\dag}=U^{-1}$.
A quantum state $|x\rangle$ $\in$ $\C^{2^n}$ is regarded as the normalized vector.


\section{Quantum algorithms for state preparation}\label{sec_main}
In this section, we introduce two types of methods for state preparation. One deterministic quantum algorithm has been proposed by Grover and Rudolph \cite{grover2002creating}.
The more general version has been proposed by Kaye and Mosca, which demands the condition probabilities are easy to compute \cite{kaye2004quantum}.

However, the quantum circuit for this algorithm is not clear enough, which may lead to confusion when analyzing the runtime. In section \ref{subsec_dtm} , we provide a specific algorithm by using quantum phase estimation \cite{nielsen2002quantum}.
This method for encoding by the phase estimation is suitable to create a probabilistic quantum algorithm we put in section \ref{subsec_prb}.
In section \ref{subsec_diag}, we introduce the decomposition of the diagonal unitary operator, which are shown in the circuits of quantum phase estimation.

Given a vector $x=(x_0e^{\bm i\theta_0},\dots,x_{N-1}e^{\bm i\theta_{N-1}})^{\rm T}\in\C^{N}$, $N=2^n$, our purpose is to prepare a state
\begin{align*}
|x\rangle=\sum_{i=0}^{N-1}e^{\bm i \theta_i}\frac{x_i}{||x||}|i\rangle,
\end{align*}
where $x_i\geqslant 0$ and $\theta_i\in[0,2\pi)$ for $i=0,\dots,N-1$.
\subsection{Deterministic quantum algorithms for state preparation}\label{subsec_dtm}
\begin{figure}[h]
  \centering
  \includegraphics[width=7cm]{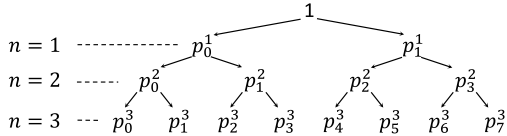}\\
  \caption{The marginal probabilities for $n=3$. }\label{fig_p}
\end{figure}

Let $x_R=(x_0,\dots,x_{N-1})^{\rm T}\in\R^{N}$ and $|x_R\rangle=\sum_{i=0}^{N-1}\frac{x_i}{||x||}|i\rangle$.
First we prepare the state $|x_R\rangle$, then add phase factors to obtain $|x\rangle$.
Suppose the probability of each component denoted by $p^n_i=x_i^2/\|x\|^2$ is known and we define the marginal probability
$p^{k}_i=\sum_{i'=2i,2i+1} p^{k+1}_{i'}$, $k=1,\dots,n-1$. Fig. \ref{fig_p} shows the case when $n=3$. For example, the first $n-1$ qubits is $|0\rangle^{\otimes {n-1}}$ with the probability $p_0^{n-1}=p^n_0+p^n_1$, and $p_2^{2}$ represents the probability in the case that the first $2$ qubits is $|10\rangle$.

We briefly describe the idea of this algorithm, then show the quantum circuit representation.
Let $|x^{k}_R\rangle=\sum_{i=0}^{2^k-1}\sqrt{p^{k}_i}|i\rangle$.
For given $|x^{k}_R\rangle$, the basic idea is to obtain $|x^{k+1}_R\rangle$ by an iteration method until $k=n-1$, that is $|x^{n}_R\rangle=|x_R\rangle$. We use quantum phase estimation algorithm to acquire the contact between $|x^{k}_R\rangle$ and $|x^{k+1}_R\rangle$ in the iteration method. Lastly we apply a diagonal operator containing phase factors to $|x_R\rangle$ to prepare $|x\rangle$.

We suppose all $x_i\neq 0$ for convenience and
remove this restriction later. The iteration procedure can be stated as follows.

{\bf Step 1.} Prepare initial state $|x_R^{1}\rangle$.
\begin{align*}
|x^{1}_R\rangle=R_Y(2\alpha^{0}_0)|0\rangle,
\end{align*}
where $\alpha^{0}_0=\arccos\sqrt{p^1_0}$. Thus, $|x^{1}_R\rangle=\sqrt{p^1_0}|0\rangle+\sqrt{p^1_1}|1\rangle$.

{\bf Step 2.} Prepare the state $|x_R^{2}\rangle$.

{\bf Step 2.1.} Define a diagonal operator $U^{1}_\alpha={\rm diag}\{e^{\bm i \alpha^{1}_0},e^{\bm i \alpha^{1}_{1}}\}$, where $\alpha^{1}_0=\arccos\sqrt{p^{2}_{0}/p^1_0}$ and $\alpha^{1}_1=\arccos\sqrt{p^{2}_{2}/p^1_1}$. The eigenvectors of $U^{1}_\alpha$ are computational basis.
Then use phase estimation algorithm to estimate $\alpha^{1}_0$, $\alpha^{1}_{1}$ by the input $|x_R^{1}\rangle$.

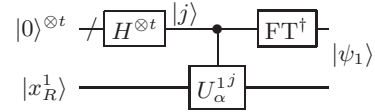
\begin{figure}[h]
\centering
\hspace{1em}\Qcircuit @C=0.5em @R=0.8em {
\lstick{|0\rangle^{\otimes t}} &\qw{/} &\gate{H^{\otimes t}} &\qw{\begin{matrix}&|j\rangle\\
\hspace{1em}\end{matrix}} &\ctrl{1} &\gate{{\rm FT^\dag}} &\qw\\
\lstick{|x_R^{1}\rangle} &\qw &\qw &\qw &\gate{{U^{1}_\alpha}^j} &\qw&\qw}
\hspace{2em}\Qcircuit @C=1em @R=.7em {
\lstick{}\\
\lstick{}
\inputgroup{1}{2}{1.1em}{|\psi_1\rangle}
}\caption{Quantum phase estimation.}\label{fig_pea}
\end{figure}

This procedure needs two registers. The first register contains $t$ qubits in the state $|0\rangle^{\otimes t}$ and the second register has a qubit  with the state $|x_R^{1}\rangle$. Use phase estimation algorithm to estimate the eigenvalues of $U^{1}_\alpha$. The circuit is represented in Figure \ref{fig_pea}.
First suppose $\alpha^{1}_l/2\pi$ is represented in binary form with $t$ bits precisely as ideal case, then analyse the general case. The final ideal state in this step is
\begin{align*}
|\psi_1\rangle=\sqrt{p^1_0}|y^{1}_0\rangle|0\rangle+\sqrt{p^1_1}|y^{1}_1\rangle|1\rangle,
\end{align*}
where $y^{1}_l\in[0,2^t-1]$ and
$y^{1}_l\pi /2^{t-1}=\alpha^{1}_l$,$l=0,1$.

\begin{figure}[h]
\centering
\hspace{1em}\Qcircuit @C=0.8em @R=1em {
\lstick{}&\ctrl{4} &\qw &\qw&\qw&\qw &\qw &\qw\\
\lstick{}&\qw &\ctrl{3} &\qw&\qw&\qw &\qw &\qw\\
\lstick{}& & &&\cdots & \\
\lstick{}&\qw &\qw &\qw&\qw&\qw &\ctrl{1} &\qw \inputgroup{1}{4}{2em}{|y^{1}_l\rangle}\\
\lstick{|0\rangle} &\gate{R_Y(2\pi)} &\gate{R_Y(\pi)}&\qw &\cdots &&\gate{R_Y(\frac{\pi}{2^{t-2}})} &\qw
}\caption{The transformation $R_Y(y^{1}_l\pi/2^{t-2})$.}\label{fig_R_Yt-2}
\end{figure}
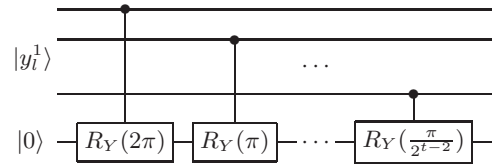

{\bf Step 2.2.} Introduce an ancilla qubit in the state $|0\rangle$ as the third register and apply $R_Y(2\alpha_l^{1})$. Due to $2\alpha_l^{1}=y_l^{1}\pi/2^{t-2}$, we can apply this rotation onto $|0\rangle$ conditioned on the first register.
The quantum circuit is shown in Figure \ref{fig_R_Yt-2}.

\begin{figure}[h]
\centering
\hspace{1em}\Qcircuit @C=0.5em @R=0.8em {
&&\qw{/} &\gate{{\rm FT}} &\qw &\ctrl{1} &\gate{H^{\otimes t}} &\qw\\
&&\qw &\qw &\qw &\gate{{U^{1}_\alpha}^{\dag j}} &\qw&\qw\\
&&\qw &\qw &\qw &\qw &\qw&\qw
\inputgroupv{1}{3}{0.01em}{2.1em}{|\psi_2\rangle}
}
\hspace{3em}\Qcircuit @C=1em @R=.7em {
\lstick{|0\rangle^{\otimes t}}\\
\lstick{}\\
\lstick{}
\inputgroup{2}{3}{2.4em}{\hspace{-0.6em}|x_R^{2}\rangle}
}\caption{Quantum phase estimation.}\label{fig_inverpea}
\end{figure}
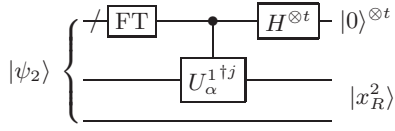

The final state in this step is
\begin{footnotesize}
\begin{align*}
|\psi_2\rangle=&\sqrt{p^1_0}|y^{1}_0\rangle|0\rangle (R_Y(2\alpha_0^{1})|0\rangle)+\sqrt{p^1_1}|y^{1}_1\rangle|1\rangle(R_Y(2\alpha_1^{1})|0\rangle),\\
=&\sqrt{p^2_0}|y^{1}_0\rangle|00\rangle+\sqrt{p^2_1}|y^{1}_0\rangle|01\rangle+\sqrt{p^2_2}|y^{1}_1\rangle|10\rangle+\sqrt{p^2_3}|y^{1}_1\rangle|11\rangle.
\end{align*}
\end{footnotesize}
{\bf Step 2.3.} Use the uncomputation technique to erase $|y_l^{1}\rangle$ and get $|x^{2}_R\rangle$. We can apply the inverse of the phase estimate transformation in the first two registers. The circuit is represented as Figure \ref{fig_inverpea}.

\begin{figure}[h]
\centering\scalebox{0.81}[0.81]{
\hspace{1em}\Qcircuit @C=0.5em @R=0.8em {
\lstick{|0\rangle^{\otimes t}} &\qw{/} &\gate{H^{\otimes t}} &\qw{\begin{matrix}|j\rangle\\
\hspace{1em}\end{matrix}} &\ctrl{1} &\gate{{\rm FT^\dag}} &\ctrl{2} &\qw &\gate{{\rm FT}} &\qw &\ctrl{1} &\gate{H^{\otimes t}} &\qw &&&|0\rangle^{\otimes t}\\
\lstick{|0\rangle} &\qw &\gate{R_Y(2\alpha^{0}_0)} &\qw &\gate{{U^{1}_\alpha}^j} &\qw &\qw &\qw &\qw&\qw &\gate{{U^{1}_\alpha}^{\dag j}} &\qw &\qw\\
\lstick{|0\rangle} &\qw &\qw &\qw &\qw &\qw &\gate{R_Y(\frac{y_i\pi}{2^{t-2}})} &\qw&\qw &\qw &\qw &\qw&\qw&\qw
}
\hspace{1.4em}\Qcircuit @C=1em @R=.7em {
\lstick{}\\
\lstick{}\\
\lstick{}
\inputgroup{2}{3}{3.2em}{|x_R^{2}\rangle}
}}
\caption{Quantum circuit to prepare $|x_R^2\rangle$.}\label{fig_deter_whole}
\end{figure}
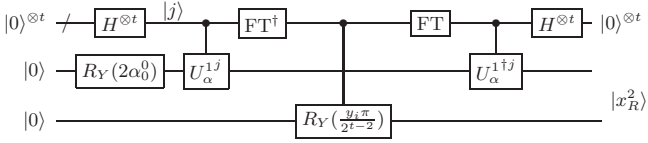

Combining step 2.1,2.2 and 2.3, we get the whole quantum circuit representation (Figure \ref{fig_deter_whole}).

{\bf Step 3.} Take the step 2 as a submodule to obtain $|x_R\rangle$.
Specifically, define a diagonal operator $U^{k}_\alpha={\rm diag}\{e^{\bm i \alpha^{k}_0},\dots,e^{\bm i \alpha^{k}_{2^k-1}}\}$, where $\alpha^{k}_j=\sqrt{p^{k+1}_{2j}/p^k_j}$.
If there exists $x_{i}=0$ such that $p^k_{i'}=0$, additionally define $\alpha^k_{i'}=0$.
Thus, the restriction $x_i\neq 0$ can be removed.
Use phase estimation algorithm to estimate $\alpha^{k}_j$, $j=0,\dots,2^k-1$ by the input $|x^{k}_R\rangle$ and continue the remaining part of step 2 to get the output state $|x^{k+1}_R\rangle$. Repeat the step 2 until $k=n-1$, that is $|x^{i+1}_R\rangle=|x_R\rangle$.

Note that $4\alpha_j^{i}\in[0,2\pi]$, then we can always estimate $2\alpha_j^{i}\in[0,\pi]\subset[0,2\pi)$ by adjust $U_\alpha^{i},R_Y(\frac{y_i\pi}{2^{t-2}})$ to $U_{2\alpha}^{i},R_Y(\frac{y_i\pi}{2^{t-1}})$. If all $\alpha_j^{i}\neq\pi/2$, we can estimate $4\alpha_j^{i}$ in similar way.

{\bf Step 4.} Add the phase factors to obtain $|x\rangle$.
Define $U_\theta=diag\{e^{\bm i\theta_0},\dots,e^{\bm i\theta_{2^n-1}}\}$, then $|x\rangle=U_\theta|x_R\rangle$.

In this algorithm, it is necessary to estimate the exact values of eigenvalues in the phase estimation algorithm.
In the general case that $\alpha^{k}_j/2\pi$ cannot be represented in binary form with $t$ bits precisely, we need to find an approximate diagonal operator $U^{i}_{\alpha^{(t)}}={\rm diag}\{e^{\bm i \alpha^{i(t)}_0},\dots,e^{\bm i \alpha^{i(t)}_{2^i-1}}\}$, where $\alpha^{i(t)}_j$ is the approximation of $\alpha^{i}_j$ and $\alpha^{i(t)}_j/2\pi$ is represented in binary form with $t$ bits precisely.
The approximate operator affect the fidelity of $|x\rangle$ we discuss in section \ref{sec_discuss}.
\subsection{Probabilistic quantum algorithms for state preparation}\label{subsec_prb}

In this section, we introduce a probabilistic quantum algorithm for state preparation. The basic framework is still the phase estimation algorithm.
Similarly, we limit $x_i\neq 0$ for convenience, and remove this restriction later.
Let $\alpha_i=\arccos (x_i/\max\{|x_i|\})\in[0,\pi/2)$.

\begin{figure}[h]
\centering\scalebox{0.8}[0.8]{
\hspace{1em}\Qcircuit @C=0.8em @R=1em {
\lstick{|0\rangle^{\otimes t}} &\qw{/} &\gate{H^{\otimes t}} &\qw{\begin{matrix}|j\rangle\\
\hspace{1em}\end{matrix}} &\ctrl{1} &\gate{{\rm FT^\dag}} \barrier[-2em]{1} &\ctrl{2}\barrier[-1.3em]{2} &\gate{{\rm FT}}&\ctrl{1}&\gate{H^{\otimes t}}&\qw &\qw&&\hspace{.8em}|0\rangle^{\otimes t}\\
\lstick{|0\rangle^{\otimes n}} &\qw{/} &\gate{H^{\otimes n}} &\qw &\gate{U_1^j} &\qw{\begin{matrix}\\\hspace{2.9em}(1)\end{matrix}} &\qw{\begin{matrix}\\\hspace{3.7em}(2)\end{matrix}} &\qw&\gate{U_1^{\dag j}}&\qw{\begin{matrix}\\\hspace{2em}(3)\end{matrix}}\barrier[-1.3em]{1}&\gate{U_2} &\qw&&|x\rangle\\
\lstick{|0\rangle} &\qw &\qw &\qw &\qw &\qw &\gate{R_Y(\frac{y_i\pi}{2^t})} &\qw&\qw&\qw&\meter &\cw &&|0\rangle
}}
\caption{The whole circuit.}\label{fig_the whole circuit}
\end{figure}
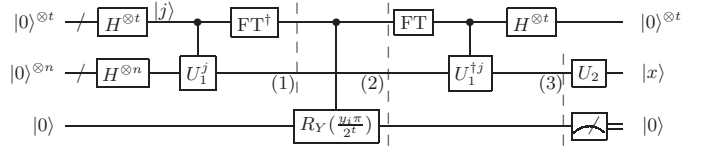

We give the whole process in Figure \ref{fig_the whole circuit}.
The unitary $U_1=diag\{e^{\bm i4\alpha_0},\dots,e^{\bm i4\alpha_{N-1}}\}$ and $U_2=diag\{e^{\bm i\theta_0},\dots,e^{\bm i\theta_{N-1}}\}$.
The controlled $R_Y(y_i\pi/2^{t})$ gate is several $R_Y(\pi/2^i)$ gates conditioned on the $i-$th qubit in the first register.

{\bf Step 1}. Prepare the state $|x_r\rangle\triangleq\sum_{i=0}^{N-1}\frac{x_i}{||x||}|i\rangle$. This notation is for distinguishing $|x_R\rangle$ in the deterministic algorithm. First suppose $4\alpha_i/2\pi$ can be precisely represented in binary form with $t$ bits.
In Figure \ref{fig_the whole circuit}, the state $(1)$ is
\begin{align*}
\frac{1}{\sqrt{2^n}}\sum_{i=0}^{2^n-1}|y^i\rangle|i\rangle,
\end{align*}
where $| y^i\rangle=|y_1^i\dots  y_t^i\rangle$ is an estimate of $4\alpha_i$.
Then we introduce an auxiliary qubit and the controlled $R_Y(y_i\pi/2^{t})$ gate. The meaning of this rotation operator can refer to Figure \ref{fig_R_Yt-2}.
The state $(2)$ is
\begin{align*}
\frac{1}{\sqrt{2^n}}\sum_{i=0}^{2^n-1}|y^i\rangle|i\rangle(\cos\alpha_i|0\rangle+\sin\alpha_i|1\rangle).
\end{align*}
Then use the uncomputation technique to erase $|y^i\rangle$ to obtain the state $(3)$
\begin{align*}
\frac{1}{\sqrt{2^n}}\sum_{i=0}^{2^n-1}|i\rangle(\cos\alpha_i|0\rangle+\sin\alpha_i|1\rangle).
\end{align*}
Next measure the auxiliary qubit. With the probability $\frac{\|x\|^2}{2^n\max_i x_i^2}$ the measure result is the state $|0\rangle$ and get the state
\begin{align}\label{eq_xr}
|x_r\rangle=\sum_{i=0}^{2^n-1}\frac{\cos\alpha_i}{\sqrt{\sum_j\cos^2\alpha_j}}|i\rangle=\sum_{i=0}^{N-1}\frac{x_i}{||x||}|i\rangle.
\end{align}

{\bf Step 2}. Prepare the state $|x\rangle$ by apply the diagonal operator $U_2$ to $|x_r\rangle$.

This algorithm is similar to the deterministic algorithm.
We need to find an approximate diagonal operator denoted by $U_1^{(t)}$ to replace $U_1$. The success probability and fidelity of preparing $|x\rangle$ are put in section \ref{sec_discuss}.

From these two algorithms, we find that except the runtime of $U^{i}_{\alpha^{(t)}}, U_1$ and $U_\theta(U_2)$ the algorithm is efficient due to $t=\mathcal O(n)$, which means the key to solve this state preparation problem is to find a suitable way of decomposing the diagonal unitary operators. Detailed discussion is presented in the section \ref{subsec_diag}.

\subsection{Diagonal unitary operator decomposition}\label{subsec_diag}

In this section, we take the decomposition of $U_1$ as an example. The general decomposition is presented in the form of the theorem.

Suppose we have acquired the accurate values of the whole $\alpha_i$. Let $\beta_i=4\alpha_i\in[0,2\pi)$.
We constantly divide the interval $[0,2\pi)$ uniformly and use $m$ to denote the number of divisions. We define
$\mathcal I_m=\{0,\frac{\pi}{2^{m-1}},\frac{2\pi}{2^{m-1}},\cdots,2\pi-\frac{\pi}{2^{m-1}}\}\subset[0,2\pi)$
and let $\beta^{(m)}_i=\max_{\beta}\{\beta\leq\beta_i, \beta\in\mathcal I_m\}$,
 that is the closed value to $\beta_i$. For example, for $m=1$ we limit the value of all the $\beta_i$ to $0$ or $\pi$ and for $m=2$ we limit  the value of $\beta_i$ to $0,\frac{\pi}{2},\pi$ or $\frac{3\pi}{2}$.

Define
\begin{align*}
U_{\beta^{(m)}}=diag\{e^{\bm i\beta_0^{(m)}},\dots,e^{\bm i\beta_{2^n-1}^{(m)}}\},
\end{align*}
and use $U_{\beta^{(m)}}$ to replace the transformation $U_1$.
Note that for all $m$ even if $x_i=0$, we have $\beta^{(m)}_i\neq2\pi$, then the restriction about $x_i$ in section \ref{subsec_prb} can be ignored.

We introduce some notations. For a quantum system with $n$ qubits, denote by $U_i$ the operator $U$ acting to the $i$-th qubit.
Generally, let $Z^{(\pm l)}=\begin{bmatrix}
1 &0\\
0 &e^{\pm 2\pi\bm i/2^l}
\end{bmatrix}$, $l\in \Z^+$, then we have
$Z^{(\pm 1)}=Z$, $Z^{(2)}=S$ and $Z^{(3)}=T$.
We denote by $C^k Z^{(l)}$ the controlled-$Z^{(l)}$ operator between the $k$ connected qubits and specifically denote by $C^k Z_{i_1,\dots,i_k}$ the operator acting to $k$ qubits, whose ordinals are $i_1,\dots,i_k$.  And we write $Z^{(l)}_i=C^1Z^{(l)}_i$.
Let $G^{(l)}$ be the group with multiplication operations generated by the set $\{I_i, C^k Z^{(l)}\}$, $i=1,\dots,n$, $k=1,\dots,n$.

When the number of division $m=1$, this problem can be converted to the construction of the  hypergraph states \cite{rossi2013quantum} and be applied in the quantum neuron \cite{tacchino2018artificial}.
We explain the decomposition of $U_{\beta^{(m)}}$ in this case.

\begin{lemma}\label{lem_m=1}
Suppose the values of all $\beta_i^{(1)}$ are known in order.
Then for $m=1$,
\begin{align}
U_{\beta^{(1)}}\in G^{(1)}.
\end{align}

\proof
Each diagonal element of $U_{\beta^{(m)}}$ is $\pm 1$,
\begin{align*}
U_{\beta^{(1)}}=\sum_{i=0}^{2^n-1}e^{\bm i \beta_i^{(1)}}|i\rangle\langle i|.
\end{align*}
We express the order $i$ in binary numbers for convenience.
First, choose the orders $I_1^{(1)}$
such that the corresponding coefficient $e^{\bm i \beta_i^{(1)}}=-1$ and the order $i$ has only one digit $1$, that is $|i\rangle=|0\dots01_{j_i}0\dots\rangle$.
Apply all the local $Z_{j_i}$ gates to $U_{\beta^{(1)}}$ to get a new diagonal unitary matrix $U_{\beta_1^{(1)}}$,
\begin{align}
U_{\beta_1^{(1)}}=\otimes_{i\in I_1^{(1)}}Z_{j_i}\otimes I\cdot U_{\beta^{(1)}}=\sum_{i=0}^{2^n-1}e^{\bm i \beta_{1i}^{(1)}}|i\rangle\langle i|.
\end{align}
We can see for $i\in I_1^{(1)}$, $e^{\bm i \beta_{1i}^{(1)}}=1$.
Secondly, choose the orders $I_2^{(1)}$
such that the corresponding coefficient $e^{\bm i \beta_{1i}^{(1)}}=-1$ and the order $i$ has only $2$ digits $1$, that is $|i\rangle=|0\dots01_{j_{1i}}0\dots01_{j_{2i}}0\dots\rangle$.
Apply all the $C^2Z_{j_{1i}j_{2i}}$ gates to $U_{\beta_1^{(1)}}$ to get
\begin{align}
U_{\beta_2^{(1)}}=\otimes_{i\in I_2^{(1)}}C^2Z_{j_{1i}j_{2i}}\otimes I\cdot U_{\beta_1^{(1)}}=\sum_{i=0}^{2^n-1}e^{\bm i \beta_{2i}^{(1)}}|i\rangle\langle i|.
\end{align}
For $i\in I_1^{(1)}\cup I_2^{(1)}$, $e^{\bm i \beta_{2i}^{(1)}}=1$.
In general, according to the first two steps, we can get the matrix $U_{\beta_k^{(1)}}$ in step $k$, $k=3,\dots,n$.
In the final step, for $i\in I_1^{(1)}\cup\cdots\cup I_n^{(1)}$, $e^{\bm i \beta_{ni}^{(1)}}=1$, which means $U_{\beta_n^{(1)}}=I$.
\begin{footnotesize}
\begin{align}
I=&C^nZ_{1\dots n}\cdots(\otimes_{i\in I_2^{(1)}}C^2Z_{j_{1i}j_{2i}}\otimes I')\cdot (\otimes_{i\in I_1^{(1)}}Z_{j_i}\otimes I'') U_{\beta^{(1)}}\notag\\
\triangleq& C^n Z\cdots C^1Z\cdot U_{\beta^{(1)}}.
\end{align}
\end{footnotesize}
Note that the inverse of $C^k Z$ is itself. Thus,
\begin{align}
U_{\beta^{(1)}}=C^1 Z\cdots C^nZ.
\end{align}
\eproof
\end{lemma}

From Lemma \ref{lem_m=1}, we know that the runtime of general decompositions of $U_{\beta^{(1)}}$ for the state preparation problems is exponential even if $m=1$. Further when the classical data is the binary strings set,  this amplitude encoding method is  still not efficient.

Our goal is to generalize this method for any $m$.
Note that $G^{(l)}\subset G^{(l+1)}$. For any $m\geq 2$, we apply some unitary transformations $C^k Z^{(m)}\in G^{(m)}$ to $U_{\beta^{(m)}}$ and obtain a matrix denoted by $U_{\beta^{(m-1)}}\in G^{(m-1)}$.
Repeat this procedure to get the decomposition of $U_{\beta^{(m)}}$.

\begin{theorem}\label{thm_m}
Suppose the values of all $\beta_i^{(m)}$ are known in order, then
\begin{align}
U_{\beta^{(m)}}\in G^{(m)}.
\end{align}

\proof
Due to $\beta_i^{(m)}\in\mathcal I_m$, each diagonal element of $U_{\beta^{(m)}}$ is $e^{\bm i\beta_i^{(m)}}=e^{\bm i \frac{p_i\pi}{2^{m-1}}}, p_i=0,\dots,2^m-1$.
In the first step, choose the orders $I_1^{(m)}$
such that the corresponding coefficient $e^{\bm i \beta_i^{(m)}}\notin \mathcal I_{m-1}$ and the order $i$ has only one digit $1$, that is $|i\rangle=|0\dots01_{j_i}0\dots\rangle$.
Apply all the local $Z^{(m)}_{j_i}$ gates to $U_{\beta^{(m)}}$ to get a new diagonal unitary matrix $U_{\beta_1^{(m)}}$,
\begin{align}
U_{\beta_1^{(m)}}=\otimes_{i\in I_1^{(m)}}Z^{(m)}_{j_i}\otimes I\cdot U_{\beta^{(m)}}=\sum_{i=0}^{2^n-1}e^{\bm i \beta_{1i}^{(m)}}|i\rangle\langle i|.
\end{align}
For $i\in I_1^{(m)}$, $\beta_{1i}^{(m)}\in\mathcal I_{m-1}$.
Secondly, choose the orders $I_2^{(m)}$
such that the corresponding coefficient $e^{\bm i \beta_{1i}^{(m)}}\notin \mathcal I_{m-1}$ and the order $i$ has only $2$ digits $1$, that is $|i\rangle=|0\dots01_{j_{1i}}0\dots01_{j_{2i}}0\dots\rangle$.
Apply all the $C^2Z^{(m)}_{j_{1i}j_{2i}}$ gates to $U_{\beta_1^{(m)}}$ to get
\begin{align}
U_{\beta_2^{(m)}}=\otimes_{i\in I_2^{(m)}}C^2Z^{(m)}_{j_{1i}j_{2i}}\otimes I\cdot U_{\beta_1^{(m)}}=\sum_{i=0}^{2^n-1}e^{\bm i \beta_{2i}^{(m)}}|i\rangle\langle i|.
\end{align}
For $i\in I_1^{(m)}\cup I_2^{(m)}$, $\beta_{2i}^{(m)}\in\mathcal I_{m-1}$.
In general, similar to the first two steps, we can get the matrix $U_{\beta_k^{(m)}}$ in step $k$, $k=3,\dots,n$.
In the final step, for $i\in I_1^{(m)}\cup\cdots\cup I_n^{(m)}$, we have $e^{\bm i \beta_{ni}^{(m)}}\in\mathcal I_{m-1}$, which means $U_{\beta_n^{(m)}}\in G^{(m-1)}$.
\begin{scriptsize}
\begin{align}
&U_{\beta_n^{(m)}}\notag\\
=&C^nZ^{(m)}_{1\dots n}\cdots(\otimes_{i\in I_2^{(m)}}C^2Z^{(m)}_{j_{1i}j_{2i}}\otimes I')\cdot (\otimes_{i\in I_1^{(m)}}Z^{(m)}_{j_i}\otimes I'') U_{\beta^{(m)}}\notag\\
\triangleq& C^n Z^{(m)}\cdots C^1Z^{(m)}\cdot U_{\beta^{(m)}}.
\end{align}
\end{scriptsize}
Note that the inverse of $C^k Z^{(m)}$ is $C^k Z^{(-m)}$. Thus,
\begin{align}
U_{\beta^{(m)}}=C^1 Z^{(-m)}\cdots C^nZ^{(-m)}\cdot U_{\beta_n^{(m)}}.
\end{align}
Continue this process for $U_{\beta_n^{(m)}}$ until the case where $m=1$, then we have
\begin{small}
\begin{align}
U_{\beta^{(m)}}=(C^1 Z^{(-m)}\cdots C^nZ^{(-m)})\cdots(C^1 Z^{(-1)}\cdots C^nZ^{(-1)}).
\end{align}
\end{small}
\eproof
\end{theorem}
The Theorem \ref{thm_m} shows that in general the gate complexity of $U_{\beta^{(m)}}$is $\mathcal O(mN)$.
In the section \ref{sec_discuss}, for the given error $\epsilon$ of the final state, the number of divisions $m(t,t\prime)$ can be regarded as $\mathcal O(n)$.
And in particular when the classical data set $X=\{x^i\}\subset \R^N(\C^N)$ is $\mathcal O(n)$-sparse, that is for all $x^i$ the number of the nonzero components is $\mathcal O(n)$, the gate complexity of the diagonal unitary operator $U_{\beta^{(m)}}$ is $\mathcal O(n^2)$.

\section{Discussions}\label{sec_discuss}
The idea for both two algorithms is first to prepare $|x_R\rangle(|x_r\rangle)$, then add the phase factors. We discuss the
fidelity for $|x_R\rangle$ and the fidelity and success probability for $|x_r\rangle$. The approximate decomposition of phase unitary transformation $U_\theta(U_2)$ only affect the fidelity of preparing $|x\rangle$ from $|x_R\rangle(|x_r\rangle)$.

\subsection{The fidelity for $|x_R\rangle$}
Since the $t$ qubits in the first register can be reused, we use $U^{i}_{\alpha^{(t)}}$ to replace $U^{i}_{\alpha}$ in all the iterative process. In this section, the relationship between $t$ and the fidelity of $|x_R\rangle$ is expounded. For convenience, we denote the approximate vectors, angles $|x^{(t)}\rangle$, $\alpha^{(t)}$ by $|\tilde x\rangle$, $\tilde\alpha$.
\begin{theorem}\label{thm_x_R}
For the deterministic algorithm, denote by $|\tilde x_R^{k}\rangle=\sum_{i=0}^{2^k-1}\sqrt{\tilde p_i^k}|i\rangle$ the final state in the $(k-1)$-th iterative process. If we choose $t=\lceil \log_2\frac{(n-1)\sqrt2\pi}{\epsilon}\rceil+1$, then the fidelity for $|x_R\rangle$ obeys the bound
\begin{align*}
||{x}_R\rangle-|\tilde x_R^{n}\rangle|\leq\epsilon.
\end{align*}
\proof
First, we discuss the error in $|x_R^{2}\rangle$.
Define $U^{1}_{\alpha^{(t)}}={\rm diag}\{e^{\bm i \tilde \alpha^{1}_0},e^{\bm i \tilde \alpha^{1}_{1}}\}$, where $\tilde \alpha^{1}_l$ is the approximation of $\alpha^{1}_l$ such that
\begin{align*}
|y_l^{1}-2^t\arccos{\sqrt{p_{2l}^2/p_l^1}/2\pi}|\leq 1,
\end{align*}
where $y_l^{1}=2^t\tilde \alpha^{1}_l/2\pi\in[0,2^t-1]\cap\Z$, $l=0,1$.
Then we have $|\tilde\alpha_l^{1}-\arccos\sqrt{p_{2l}^2/p_{2l}^1}|\leq \pi/2^{t-1}$. Thus,
\begin{align*}
||x_R^{2}\rangle-|\tilde x_R^{2}\rangle|\leq \sqrt{2}\pi/2^{t-1}.
\end{align*}

Perform the $(k-1)$-th iteration from $|\tilde x_R^{k-1}\rangle$ to prepare $|\tilde x_R^{k}\rangle$.
Let $|{x'}_R^{k}\rangle=\sum_{i=0}^{2^k-1}\sqrt{{p'}_i^k}|i\rangle$ such that $\tilde p^{k-1}_i=\sum_{i'=2i,2i+1}{p'}_{i'}^{k}$ and ${p'}_{2i}^{k}/{p'}_{2i+1}^{k}={p}_{2i}^{k}/{p}_{2i+1}^{k}$.
Then
\begin{align*}
||{x'}_R^{k}\rangle-|\tilde x_R^{k}\rangle|\leq \sqrt2\pi/2^{t-1}.
\end{align*}
and
\begin{align*}
||{x}_R^{k}\rangle-|{x'}_R^{k}\rangle|=||x_R^{k-1}\rangle-|\tilde x_R^{k-1}\rangle|\leq (k-2)\sqrt2\pi/2^{t-1}.
\end{align*}
Thus,
\begin{align*}
||{x}_R\rangle-|\tilde x_R^{n}\rangle|\leq (n-1)\sqrt2\pi/2^{t-1}.
\end{align*}
\eproof\end{theorem}
\subsection{The fidelity and success probability for $|x_r\rangle$}
\begin{theorem}\label{thm_x_r}
For the probabilistic algorithm, set $t=2n+\lceil\log_2\frac{\pi}{\epsilon}\rceil$, then with success probability greater than $\frac{\|x\|^2}{2^n\max_i x_i^2}$ the fidelity for $|x_r\rangle$ obeys the bound
\begin{align*}
||{x}_r\rangle-|\tilde x_r\rangle|\leq\epsilon.
\end{align*}
\proof
Since $\tilde\alpha_i\leq\alpha_i$, the probability to measure $|0\rangle$ is
\begin{align*}
\frac{\sum_{i=0}^{2^n-1}\cos^2\tilde\alpha_i}{2^n}\geq\frac{\sum_{i=0}^{2^n-1}\cos^2\alpha_i}{2^n}=\frac{\|x\|^2}{2^n\max_i x_i^2}.
\end{align*}
We get
\begin{align*}
|\tilde x_r\rangle=\sum_{i=0}^{2^n-1}\frac{\cos\tilde\alpha_i}{\sqrt{\sum_j\cos^2\tilde\alpha_j}}|i\rangle.
\end{align*}
By \eqref{eq_xr}, we have
\begin{scriptsize}
\begin{align*}
||x_r\rangle-|\tilde x_r\rangle|\leq&\sum_{i=0}^{2^n-1}\left|\frac{\cos\alpha_i}{\sqrt{\sum_j\cos^2\alpha_j}}-\frac{\cos\tilde\alpha_i}{\sqrt{\sum_j\cos^2\tilde\alpha_j}}\right|\\
\leq&\sum_{i=0}^{2^n-1}\left|\frac{\cos\alpha_i}{\sqrt{\sum_j\cos^2\alpha_j}}-\frac{\cos\tilde\alpha_i}{\sqrt{\sum_j\cos^2\alpha_j}}\right|\\
&+\sum_{i=0}^{2^n-1}\left|\frac{\cos\tilde\alpha_i}{\sqrt{\sum_j\cos^2\alpha_j}}-\frac{\cos\tilde\alpha_i}{\sqrt{\sum_j\cos^2\tilde\alpha_j}}\right|.
\end{align*}
\end{scriptsize}
Since $|\alpha_i-\tilde\alpha_i|<\pi/2^{t+1}\triangleq\delta$, we have
\begin{align}\label{eq_cos_m}
|\cos\alpha_i-\cos\tilde\alpha_i|<\delta.
\end{align}
Note that $|x_r\rangle\neq0$, which means $\sum_i\cos^2\alpha_j\geq1$ and we have
\begin{align}\label{eq_1/cos}
\left|\frac{1}{\sqrt{\sum_j\cos^2\alpha_j}}-\frac{1}{\sqrt{\sum_j\cos^2\tilde\alpha_j}}\right|\leq2^{n}\delta.
\end{align}
Combining expressions \eqref{eq_cos_m}\eqref{eq_1/cos}, we obtain
\begin{align*}
||x_r\rangle-|\tilde x_r\rangle|\leq 2^{2n+1}\frac{\pi}{2^{t+1}}.
\end{align*}
Let $t=2n+\lceil\log_2\frac{\pi}{\epsilon}\rceil$, we have $||{x}_r\rangle-|\tilde x_r\rangle|\leq\epsilon$.
\eproof\end{theorem}
\subsection{The fidelity and success probability for $|x\rangle$}
Suppose $|x_R\rangle(|x_r\rangle)$ is prepared precisely. By Setting $U_{\theta^{(t')}}$ to replace $U_\theta(U_2)$, we have $||x\rangle-|x^{(t')}\rangle|\leq2^n\pi/2^{t'-1}$. Let $\epsilon'=\epsilon/2$, then we obtain $|x\rangle-|x^{(t')}\rangle|\leq\epsilon'$ when setting $t'=n+1+\lceil\log_2\frac{\pi}{\epsilon'}\rceil$.
Combine Theorem \ref{thm_x_R} and Theorem \ref{thm_x_r}, then we have the result
\begin{theorem}
For the deterministic algorithm, set $t=\lceil \log_2\frac{2(n-1)\sqrt2\pi}{\epsilon}\rceil+1$, $t'=n+1+\lceil\log_2\frac{2\pi}{\epsilon}\rceil$, then the fidelity for $|x\rangle$ obeys the bound
\begin{align*}
||x\rangle-|\tilde x\rangle|\leq\epsilon.
\end{align*}
\end{theorem}
\begin{theorem}
For the probabilistic algorithm, set $t=2n+\lceil\log_2\frac{2\pi}{\epsilon}\rceil$, $t'=n+1+\lceil\log_2\frac{2\pi}{\epsilon}\rceil$, then with success probability greater than $\frac{\|x\|^2}{2^n\max_i x_i^2}$ the fidelity for $|x\rangle$ obeys the bound
\begin{align*}
||x\rangle-|\tilde x\rangle|\leq\epsilon.
\end{align*}
\end{theorem}
\section{Conclusion}\label{sec_conclusion}

We introduce two algorithms of amplitude encoding for state preparation problems.
In this method, for a vector $x\in \C^N$ we need $\log_2 N$ qubits to construct the corresponding quantum state $|x\rangle$ with exponential depth.
We analyse the fidelity for the deterministic algorithm and the probabilistic algorithm and discuss the success probability in the probabilistic algorithm. This success probability is decided by the data structure.

We turn the state preparation problem to the decomposition problem of the diagonal unitary operators. The runtime in these algorithms whose corresponding circuits are limited to the width $\mathcal O(n)$ is exponential to $n$ generally. However, in certain cases, the decomposition of diagonal unitary transformations naturally only contains polynomial layers; e.g., the classical data set is $O(n)$-sparse. In other words, the quantum states whose corresponding diagonal unitary operators have polynomial decompositions can be prepared effectively.

\section*{acknowledgements}

This work was supported by the National Key Research and Development
Program of China (Grant No. 2016YFA0301700) and the Anhui Initiative in
Quantum Information Technologies (Grant No. AHY080000).

\bibliographystyle{unsrt}
{\small
 \bibliography{123}
}

%
%
%

%
%
%
%
%
%
%
%

\end{document}